\documentclass[12pt]{article}
\usepackage{newtxtext,newtxmath}
\usepackage{graphicx}
\usepackage[letterpaper,margin=1in]{geometry}

\renewenvironment{abstract}
	{\quotation}
	{\endquotation}

\date{}

\makeatletter
\renewcommand{\fnum@figure}{\textbf{Figure \thefigure}}
\renewcommand{\fnum@table}{\textbf{Table \thetable}}
\makeatother

\usepackage{scicite}

\usepackage{url}

\def\scititle{
	A Route to Quantum-Coherent Thermodynamics
}
\title{\bfseries \boldmath \scititle}
\author{Maurizio~Fagotti$^{1\ast}$\and
\small$^{1}$Universit\'e Paris-Saclay, CNRS,  Laboratoire de Physique Théorique et Modèles Statistiques\and
	\small$^\ast$ Email: maurizio.fagotti@universite-paris-saclay.fr\and
}

\begin{document} 

\maketitle
\begin{abstract} \bfseries \boldmath
Equilibrium thermodynamics emerges because a vast number of microscopic states become indistinguishable at the level of local observables. We start from the observation that a nonequilibrium quantum state can also be represented by an ensemble of pure states that minimizes the average energy variance. We show that varying the statistical weights within this optimal family organizes state space into “minimum-variance leaves”, which generalize the space of stationary states underlying equilibrium. We observe thermodynamic compression on the leaves, which in the simplest scenario manifests itself in the indistinguishability of states with the same energy density. We call this “leaf typicality”; at equilibrium it reduces to eigenstate thermalization or its generalizations. Numerical tests support leaf typicality on some leaves and indicate a breakdown of its stronger forms on others, suggesting a rich thermodynamic organization beyond equilibrium. These results open a route to a unified statistical mechanics of equilibrium and nonequilibrium quantum states.
\end{abstract}

\noindent
Many-body quantum systems quickly exceed the reach of fully microscopic analysis, and even when such analyses are possible, their implications are most naturally read through thermodynamic lenses. Statistical mechanics enables this by distilling microscopic complexity into a tractable set of macroscopic properties that govern accessible observables. This reduction is particularly effective at equilibrium, whereas constructing an equally systematic and efficient framework out of equilibrium has been a long-standing objective since the origins of statistical mechanics~\cite{Boltzmann2003Further}.
From the perspective of quantum information theory, ensembles of energy eigenstates are special because their states carry no energy coherence: the quantum Fisher information (QFI)~\cite{Braunstein1994Statistical,Paris2009Quantum} of any stationary state with respect to the Hamiltonian $H$ vanishes, $F_Q(\rho;H)=0$.
For nonstationary states, the QFI is nonzero, revealing an irreducibly quantum contribution to energy fluctuations that can affect their thermodynamic properties.  To simplify and systematize the description of time evolution, a common approach is to focus on subsystems or, more generally, open systems. Attention then turns to dynamical mechanisms, such as dephasing~\cite{Barthel2008Dephasing,Reimann2008Foundation,Linden2009Quantum,deOliveira2018Equilibration}, that suppress the observable effects of energy coherence and can lead to a quasi-stationary description.
This has encouraged an emphasis on  how relaxation to equilibrium occurs, and what notion of equilibration is appropriate~\cite{Polkovnikov2011Colloquium,Eisert2015Quantum,Rigol2007Relaxation,Essler2016Quench}. 
The appeal of equilibration lies in its drastic reduction of complexity, which culminates in the eigenstate thermalization hypothesis (ETH)~\cite{Deutsch1991Quantum,Srednicki1994Chaos,Rigol2008Thermalization,DAlessio2016From, Kuwahara2020Eigenstate}:  
local observables depend essentially only on the energy density of the eigenstate.
From this point of view, nonequilibrium thermodynamics  becomes a multifaceted venture to extend equilibrium concepts beyond their established domain.

There is, however, an important complementary perspective.
Canonical typicality establishes the local equivalence of almost all pure states within an equilibrium energy shell~\cite{Goldstein2006Canonical,Popescu2006Entanglement}, while dynamical typicality extends this to the time evolution of states subject to prescribed constraints~\cite{Reimann2007Typicality,Bartsch2009Dynamical,Reimann2018Dynamical}. These results reveal a remarkable insensitivity to microscopic details within the prescribed ensemble even for nonstationary states.

Although well established in many respects, the theory now faces new challenges posed by next-generation experiments.
Quantum platforms based on ultracold atoms, trapped ions, and Rydberg-atom arrays maintain coherent quantum dynamics for long times and can probe entire systems as well as subsystems comprising an appreciable fraction of the full system~\cite{Kaufman2016Quantum,Bernien2017Probing,Brydges2019Probing}. In such regimes, energy coherence cannot always be treated as a perturbation, motivating the development of thermodynamic descriptions that incorporate its structure nonperturbatively.

Here we pursue this ultimate goal by providing a crucial missing ingredient: an organization of state space compatible with thermodynamic compression, pointing the way towards a comprehensive theory of nonequilibrium thermodynamics, a quantum-coherent thermodynamics.

An analogy useful to understand our perspective is provided by quantum measurement theory. A projective measurement of a nondegenerate observable $H$ with eigenstates ${|\varphi_i\rangle}$, when its outcome is ignored, yields
\begin{equation}\label{eq:rhop}
\rho[{p}]=\sum_i p_i|\varphi_i\rangle\langle\varphi_i|\,,
\end{equation}
where $p_i$ are the outcome probabilities. When $H$ is the Hamiltonian, this class consists precisely of stationary states and includes the equilibrium statistical ensembles. The extension from projective to generalized measurements~\cite{Davies1970An} suggests a parallel step: relaxing the orthogonality of the ensemble in Equation~\ref{eq:rhop}.
Specifically, we propose a common principle for representing (almost) arbitrary quantum states by generally nonorthogonal pure-state ensembles, so that the thermodynamic compression of local information familiar from the
energy eigenbasis extends consistently across state space.

\section*{Results}
\subsection*{Minimum-variance foliation}
Relaxing the assumption that ensemble states are stationary requires a shift of principle. 
Our starting point is the equivalence of statistical ensembles at equilibrium~\cite{Lima1972Equivalence}, for which controlling fluctuations is a key implicit requirement~\cite{Tasaki2018On,Kuwahara2020Gaussian,BrandaoCramer2015Equivalence}. Seeking analogous equivalence classes away from equilibrium, we promote this requirement to a principle of \emph{minimal coherent energy fluctuations}.
To this end, we single out families of pure states $\{|\varphi_i\rangle\}$ (generally nonorthogonal) with the
following property:

\emph{For every choice of probabilities $\{p_i\}$, the  decomposition
$\rho=\sum_i p_i|\varphi_i\rangle\langle\varphi_i|$
minimizes the ensemble-averaged energy variance among all pure-state decompositions of~$\rho$.}
In other words, for all $\{p_i\}$ one has
\begin{equation}\label{eq:family}
\begin{aligned}
&\sum_i p_i\,\mathrm{Var}_{\varphi_i}(H)
=
\inf_{\substack{\rho=\sum_j p'_j|\psi_j\rangle\langle\psi_j|}}
\sum_j p'_j\,\mathrm{Var}_{\psi_j}(H)\\
&\mathrm{Var}_{\psi}(H):=\langle\psi|H^2|\psi\rangle-\langle\psi|H|\psi\rangle^2 .
\end{aligned}
\end{equation}
It is well known~\cite{Toth2013Extremal,Yu2013Quantum} that the minimal average energy variance (over all pure-state decompositions) is directly related to the QFI of $\rho$ with respect to $H$\footnote{If $\{|\varphi_i\rangle\}$ satisfy Equation~\ref{eq:family}, then $F_Q(\rho;H)=4\sum_i p_i \mathrm{Var}_{\varphi_i}(H)$.}. 
Less appreciated is that, by convexity, the same family remains optimal under arbitrary changes of its populations~\cite{Uhlmann2010Roofs,Regula2018Convex}.
In more mathematical terms, the minimum-variance decompositions induce a foliation of (a dense open subset of) the state space, with each leaf associated with a fixed family of pure states~\cite{Uhlmann2010Roofs}.
The optimal families selected by Equation~\ref{eq:family} are therefore our candidates for extending the thermodynamic role of energy eigenstates to nonstationary states.   

Following Yu~\cite{Yu2013Quantum}, we define an effective ``dressed Hamiltonian'' $H_\rho$ (called $Y$ in Ref.~\cite{Yu2013Quantum}) as the operator satisfying
\begin{equation}\label{eq:Hrho}
\frac{1}{2}\{H_\rho,\rho\}=\rho^{\frac{1}{2}} H \rho^{\frac{1}{2}}\, .
\end{equation}
This operator reduces to $H$ if $\rho$ commutes with $H$. In addition, we have $\mathrm{tr}(\rho H_\rho)=\mathrm{tr}(\rho H)$.
Ref.~\cite{Yu2013Quantum} showed that the optimal ensemble consists of a number of states equal to the rank of $\rho$ and the populations are directly related to the eigenvectors, $|\Psi_i\rangle$, and eigenvalues, $E_{i}$, of $H_\rho$, $H_\rho |\Psi_i\rangle=E_{i}|\Psi_i\rangle$,  as follows
\begin{equation}\label{eq:Yu}
\begin{aligned}
%&p_i=\braket{\Psi_i|\rho|\Psi_i}\\
&p_i |\varphi_i\rangle\langle\varphi_i|=\sqrt{\rho}|\Psi_i\rangle\langle\Psi_i|\sqrt{\rho}\\
&\langle\varphi_i|H|\varphi_i\rangle=E_{i}\, .
\end{aligned}
\end{equation}
As a side mathematical note, we remark that a genuine foliation requires leaves to be pairwise disjoint. This fails on the full state space: at points where $H_\rho$ is degenerate, the minimum-variance ensemble is not unique, and distinct leaves meet. Nevertheless, the leaves are disjoint on the open subset $\mathcal M_H:=\{\rho:H_\rho\text{ is nondegenerate}\}$, 
where the minimizing family of pure states is uniquely determined~\cite{Yu2013Quantum} (up to phases and permutation) and therefore labels a single leaf. Whenever we speak of a foliation we implicitly restrict to $\mathcal M_H$. For a generic choice of $H$ (in particular, after lifting accidental degeneracies), $\mathcal M_H$ is  open and dense in the space of density matrices. As a rule of thumb, if $\rho$ is thermal for some quasilocal Hamiltonian, it belongs to $\mathcal M_H$ only if its temperature is nonzero.

One of the most important properties of the foliation is that  all states on a leaf share the same symmetric logarithmic derivative (SLD) $L_\rho^{(H)}$ associated with $H$---cf.~Equation~\ref{eq:Lyapunov}. If $H$ is the Hamiltonian of a quantum spin chain with (quasi)local interactions, the SLD can be written as $L_\rho^{(H)}=2\sum_\ell \tilde h_{\rho,\ell}$ and its density $2\tilde h_{\rho}$ can be expressed in terms of the density $h_{\rho}$ of the dressed Hamiltonian $H_\rho=\sum_\ell h_{\rho,\ell}$ as   
$\tilde h_{\rho,\ell}=\frac{i}{2}(\rho^{\frac{1}{2}}h_{\rho,\ell} \rho^{-\frac{1}{2}} - \rho^{-\frac{1}{2}}h_{\rho,\ell} \rho^{\frac{1}{2}} )$. 
Under mild physical assumptions on the state, $\tilde h_{\rho}$ is quasilocal and is determined by the local properties of the state (Materials and Methods). Thus, \emph{states belonging to thermodynamically distinct leaves}, i.e. leaves with inequivalent SLD densities, \emph{cannot be locally equivalent}. This is a strong indication that leaf structure plays a central role in the thermodynamic description and directs our attention to how thermodynamic compression manifests itself within the leaves of the foliation. 

\subsection*{Leaf statistical ensembles}
From Equation~\ref{eq:Yu} it readily follows  that, if $\rho$ has full rank, any other density matrix in the (closure of the) same leaf (i.e. obtained by varying the populations while keeping the same optimal family) can be written as 
\begin{equation}
\rho'=\frac{\sqrt{\rho}R_H^{(\rho)}\sqrt{\rho}}{\mathrm{tr}[\rho R_H^{(\rho)}]},
\qquad
R_H^{(\rho)} \geq 0,\quad [R_H^{(\rho)},H_\rho]=0.
\end{equation}
We are now in a position to construct statistical ensembles that capture the typical properties of the states $|\varphi_i\rangle$.
We identify the thermodynamic entropy on the leaf with the Shannon entropy of the populations, $S_{th}^{(H)}[\rho]=-\sum_i p_i \log p_i$. This is a natural choice as it is independent of the coherent structure\footnote{Note that, instead, the von Neumann entropy mixes population uncertainty with the energy coherent structure.} encoded in the nonorthogonality of the states $|\varphi_i\rangle$ (it only depends on the coordinates on the leaf) and is consistent with some basic thermodynamic relations enjoyed also by the diagonal entropy~\cite{Polkovnikov2011Microscopic}---Materials and Methods.  
In analogy with the Gibbs--Jaynes maximum-entropy prescription~\cite{Jaynes1957Information},
we then define the \emph{minimum-variance leaf canonical ensemble} by maximizing the thermodynamic entropy, subject to normalization and fixed mean energy. Since $\{E_{i}\}$ are 
 independent of $\{p_i\}$ and within a given leaf the energy reads
$
E[\rho]=\sum_i p_i\langle\varphi_i|H|\varphi_i\rangle=
\sum_i p_i E_{ i}
$,
the maximization yields the usual Gibbs weights $p_i\propto e^{-\beta E_{i}}$. 
More generally, it is convenient to define the leaf canonical ensemble through a density matrix $\rho$ as 
\begin{equation}\label{eq:leaf-canonical}
\rho_{\beta|\mathcal L_H(\rho)}=\frac{\sqrt{\rho}\exp(-\beta H_{\rho})\sqrt{\rho}}{\mathrm{tr}[\rho \exp(-\beta  H_{\rho})]}\, ,
\end{equation}
where $\mathcal L_H(\rho)$ denotes the leaf through $\rho$.
It then follows that the maximum entropy prescription selects the leaf canonical ensemble through the reference density matrix $\rho_\bullet$ with uniform weights (equal to the inverse of the dimension of the Hilbert space), which we dub ``barycenter of the leaf''. 
Standard equilibrium is recovered as the special case where the state is restricted to the commuting leaf, $[\rho,H]=0$, which corresponds to $\rho_\bullet\propto \mathrm I$. 
We caution against assigning a physical meaning to $\beta$, as its thermodynamic role cannot be inferred from this finite-volume construction alone~\cite{Fagotti2026Temperature}.

Analogously, we define a leaf-microcanonical ensemble by replacing $\exp(-\beta H_{\rho})$ in Equation~\ref{eq:leaf-canonical} with the projector onto the chosen energy shell of $H_{\rho}$, and then normalizing.

We stress that varying the populations preserves the coherent structure encoded in the geometry of the leaf $\mathcal L_H$. We characterize the leaf’s degree of incoherence by the von Neumann entropy of its barycenter, denoted by $\mathfrak I(\mathcal L_H)$; see Equation~\ref{eq:incoherence} in Materials and Methods.

\subsubsection*{Example}
For a single spin-$\frac{1}{2}$, the minimum-variance foliation admits a simple geometric picture.
Without loss of generality we set $H=\varepsilon\sigma^z$ and represent states as points in the Bloch ball,
$\rho=\frac{1}{2}(\mathrm I+\vec r\cdot\vec\sigma)$ with $|\vec r|\leq 1$.
The leaves are the straight chords parallel to the $z$ axis: each leaf is specified by a fixed transverse component
$\vec r_\perp=\vec r-(\vec r\cdot\hat z)\hat z$, while the population parameter moves the state along $z$.
The two pure states at the endpoints of a leaf have Bloch vectors $\hat n$ and its reflection through the $xy$-plane,
$\hat n'=\hat n-2(\hat n\!\cdot\!\hat z)\hat z$.
Their barycenter is therefore
$
\rho_\bullet=\frac{1}{2}(\mathrm I+[\hat n-(\hat n\cdot\hat z)\hat z]\cdot\vec\sigma),
$
whose Bloch-vector length is $|\hat n-(\hat n\cdot\hat z)\hat z|=|\hat n\times\hat z|$.
Hence the von Neumann entropy of the barycenter is
$
S_{vN}[\rho_\bullet]=H_1(\frac{1-|\hat n\times\hat z|}{2})$, with 
$
H_1(x)=-x\log x-(1-x)\log(1-x)
$.
According to our leaf-level incoherence measure, the most coherent leaves are those with $|\hat n\times\hat z|\rightarrow 1$
(i.e., equatorial spins, $\hat n\cdot\hat z\rightarrow 0$), for which $\mathfrak I(\mathcal L_H)\rightarrow 0$ and the energy QFI is maximal.

The first mathematical complications arise in dimension~$3$ (a single spin-$1$), where, without restricting the space of states to $\mathcal M_H$, the leaves would intersect.  

\subsection*{Leaf typicality}

Unlike the commuting leaf, a generic leaf is not invariant under the time evolution generated by $H$: it is transported to another leaf while preserving the multiset of energy variances of its pure-state ensemble. This constraint is nevertheless strong and implies that, even when a leaf is considered as a whole, unitary evolution explores only a thin subset of state space (cf. Ref.~\cite{Fagotti2019On}).
In the spirit of ETH, locality can lead to a further reduction once states that are indistinguishable by local observables are identified. 
Motivated by the local structure underlying the SLD characterization of the leaves, we conjecture that the ergodic mechanism underlying ETH on the commuting leaf extends more generally (cf. Ref.~\cite{Foini2025Out}), reducing the states explored by a nonequilibrium leaf to a manifold of locally distinguishable states whose dimension remains finite in the thermodynamic limit. 
The first step is embodied in the following question. For any local observable $O$ and leaf $\mathcal L_H$, does there exist a smooth function $f_{O,\mathcal L_H}$ such that, for eigenstates $|\Psi_i\rangle$ of $H_\rho$ with $E_{i}\approx E$ (cf.~Equation~\ref{eq:Yu}),
\begin{equation}\label{eq:leaf_typicality}
\frac{\langle\Psi_i|\sqrt{\rho}O \sqrt{\rho}|\Psi_i\rangle}
{\langle\Psi_i|\rho|\Psi_i\rangle}\equiv \langle\varphi_i|O|\varphi_i\rangle
\sim f_{O,\mathcal L_H}(E)\, ,
\end{equation}
with fluctuations within each energy shell vanishing as the system size grows?
We call this property \emph{leaf typicality}\footnote{Note that leaf typicality is not equivalent to ETH for $H_\rho$: the expectation value in Equation~\ref{eq:leaf_typicality} involves $\sqrt{\rho}O\sqrt{\rho}$  and $\rho$, neither of which is quasilocal.}. As in ETH, we qualify it as strong or weak depending on whether Equation~\ref{eq:leaf_typicality} holds for every eigenstate of $H_\rho$ or for all but a vanishing fraction of them. We further use the refinement \emph{near-strong} when the number of outliers grows more slowly than any power of the Hilbert-space dimension. 
It is precisely this hypothesis that connects the finite-dimensional foliation introduced above with the thermodynamic regime, where the physically relevant information is encoded in local observables. 

We point out that this hypothesis does not follow from conventional typicality results. For example, in translation-invariant systems weak ETH follows from clustering properties~\cite{Biroli2010Effect,Mori2016Weak} (which Gibbs states  in 1D enjoy~\cite{Araki1969Gibbs}). The clustering argument, however, is not sufficient on other leaves because the thermodynamic entropy $S_{th}^{(H)}[\rho]$ can be extensively smaller than $S_{th}^{(H)}[\rho_{\beta_\rho|\mathcal L_H(\rho_\bullet)}]$, with $\beta_\rho$ fixed by $\mathrm{tr}(\rho_{\beta_\rho|\mathcal L_H(\rho_\bullet)}H) = \mathrm{tr}(\rho H)$. In this situation, the inequality $\sum_i p_i(\langle\varphi_i|O|\varphi_i\rangle-\mathrm{tr}(\rho O))^2\leq \mathrm{Var}_\rho(O)$, applied to spatial averages of local observables, bounds the total probability of outliers, without establishing that their unweighted fraction vanishes.

\begin{figure}[t]
\centering
\includegraphics[width=0.5
\textwidth]{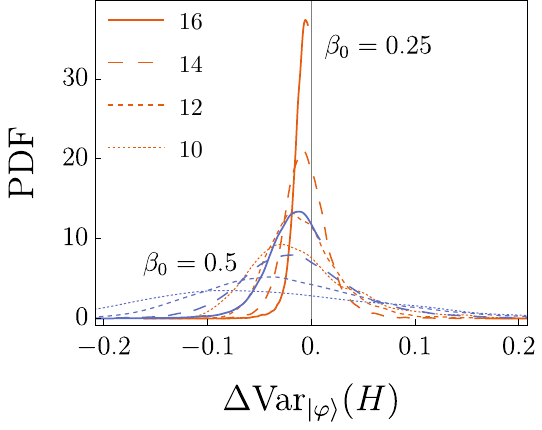}
\caption{
Unweighted distributions of the deviations of the pure-state energy variance (extensive in this system) from its local average in energy, for the chain lengths indicated in the legend. The average is estimated as one quarter of the quantum Fisher information of the isoenergetic leaf canonical ensemble through the initial state.  The Hamiltonian $H$ is as in Equation~\ref{eq:H} with parameters $\vec h=(\frac{\sqrt{5}+5}{8},\frac{1}{2},\frac{\sqrt{5}}{2})$ and $D=\frac{\pi}{20}$. The state is thermal $\rho_{\beta_0} \propto e^{-\beta_0 H_0}$,  where $H_0$ is as in Equation~\ref{eq:H} with $\vec h=(0,0,\frac{3}{2})$ and $D=0$. 
}\label{f:ETHtime1}
\end{figure}

To illustrate this organization in a quantum many-body system, we report a numerical investigation on \mbox{spin-$\frac{1}{2}$} chains of length $L\leq 16$ (Hilbert-space dimension $d=2^L$). 
We consider a local Hamiltonian of the form 
\begin{equation}\label{eq:H}
H=\sum\nolimits_{\ell}^L\sigma_\ell^x\sigma_{\ell+1}^x+\vec h\cdot \vec \sigma_\ell
+D\big(\sigma_\ell^z\sigma_{\ell+1}^y-\sigma_\ell^y\sigma_{\ell+1}^z\big),
\end{equation}
where $\sigma_\ell^\alpha$ acts as a Pauli matrix on site $\ell$ and as the identity elsewhere, and impose periodic boundary conditions $\sigma_{L+1}^\alpha\equiv \sigma_1^\alpha$.
We choose the couplings so that ETH signatures are already visible at those sizes for all observables supported on one site. 
We construct $\rho$ as a thermal state $\rho=\rho_{\beta_0}\propto e^{-\beta_0 H_0}$ of a different Hamiltonian $H_0$ within the same family; specifically, we take an integrable transverse-field Ising point. 
We first examine the quantity underlying the construction: the energy variance in the pure states of the optimal decomposition. Equation~\ref{eq:EVFO} in Materials and Methods expresses these variances as expectation values of an observable with quasilocal density; their concentration, shown in Figure~\ref{f:ETHtime1}, therefore provides evidence for weak leaf typicality. 

\begin{figure}[t]
\centering
\includegraphics[width=
\textwidth]{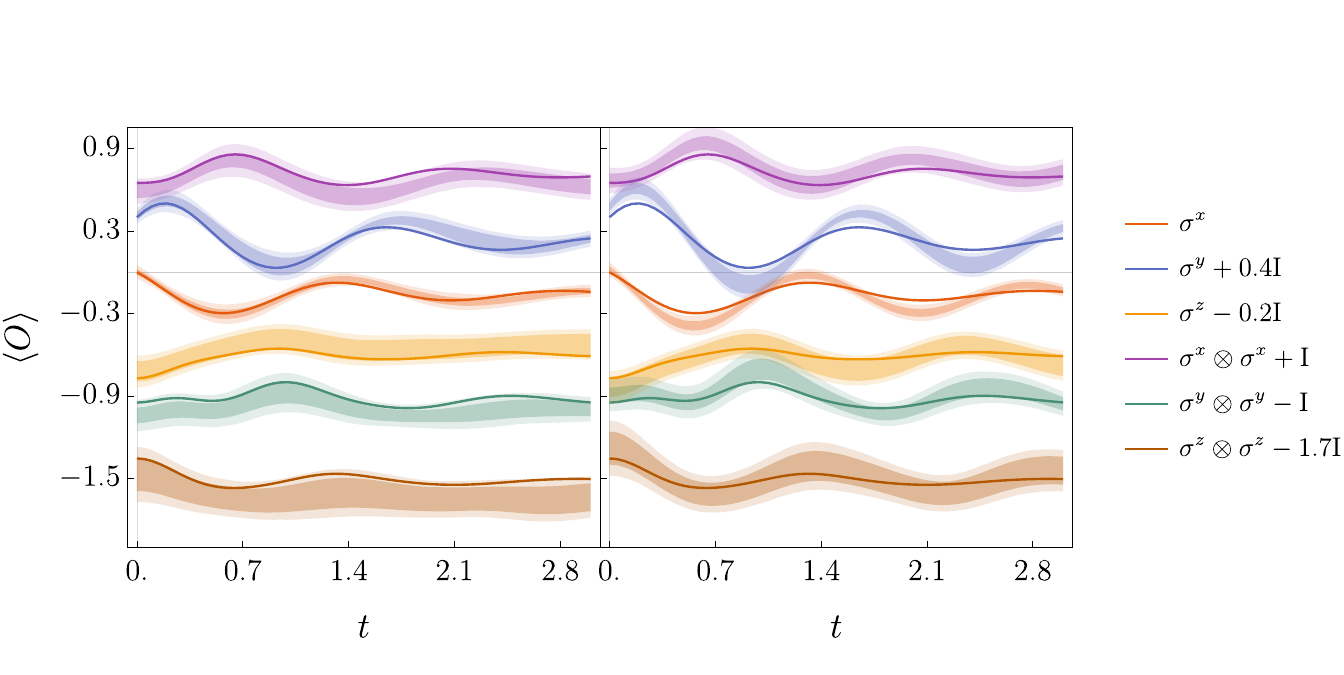}
\caption{ 
Exact dynamics under the same Hamiltonian $H$ as in Figure~\ref{f:ETHtime1} from the thermal state with $\beta_0=0.5$ (solid lines), compared to predictions obtained by evolving a representative state $|\varphi_i\rangle$ of the optimal decomposition for $L=16$.  The two dynamical panels correspond to the first and second states, respectively, if the states are ordered with respect to the energy deviation $\delta_i=|E_{i}-\langle H\rangle|/(\mathrm{Var}_{\varphi_i}(H)+\frac{1}{4}F_Q(\rho;H))^{1/2}$. Shaded bands indicate $68\%$ (dark) and $95\%$ (light) coverage intervals, centered at the representative evolutions, estimated from the outlier fraction within the $\delta$-shell defined by $\delta_i-\min_i\delta_i\leq\frac{1}{L}(\max_i\delta_i-\min_i\delta_i)$ at each given time.
}\label{f:ETHtime2}
\end{figure}

Before challenging the stronger forms of leaf typicality, we test one of its most direct consequences: within any fixed finite time window, the evolution of local expectation values should be reproduced, in the thermodynamic limit, by evolving (almost) any isoenergetic pure-state representative drawn from the optimal ensemble.
We stress that representative-state ideas are not new: e.g., it is well known that the local properties of a thermal state can be reproduced by a representative pure state~\cite{Sugiura2013Canonical} or that a generalized Gibbs ensemble of an integrable system is locally equivalent to a saddle-point state~\cite{Caux2013Time,Caux2016The}. 
Here, however, we are concerned with the thermodynamic properties of states rather than their typical behavior.
On the one hand, sequences of leaves, indexed by system size, whose SLD densities remain uniformly quasilocal allow the constraints underlying the foliation to be formulated in the thermodynamic limit~\cite{Fagotti2026Temperature}. On the other hand, the degree to which leaf typicality holds measures how strongly the energy-coherence structure fixes local properties at fixed energy density.
Figure~\ref{f:ETHtime2} shows the dynamics in a chain of $16$ spins and  illustrates how the typicality properties of a leaf are shared by all the leaves through the state at different times (with the implicit complication that time becomes a relevant parameter for the locality properties of the observables~\cite{Bravyi2006Lieb}). The close agreement is particularly revealing because, in the sense of dynamical typicality, the representative states of the optimal decomposition are  atypical. Indeed, a typical pure state drawn from a conventional random-state ensemble representing $\rho$ also approximately reproduces its energy variance and higher moments. 
In the minimum-variance ensemble, instead, the average energy variance of the representatives equals the coherent contribution $\frac{1}{4}F_Q(\rho;H)$, which accounts for about $31\%$ of the total energy variance in Figure~\ref{f:ETHtime2}.

\begin{figure}[t]
\centering
\includegraphics[width=0.7\textwidth]{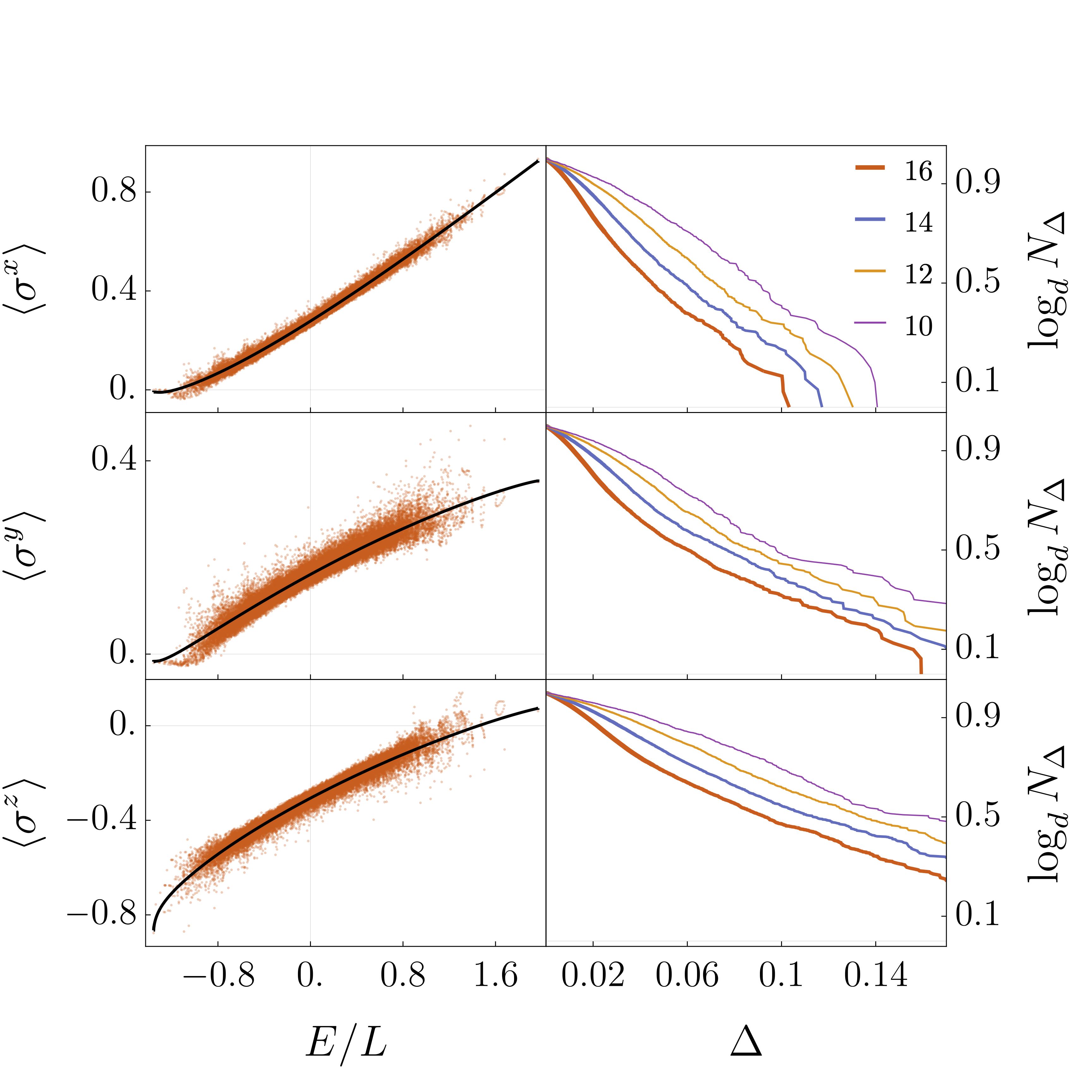}
\caption{
Indications of near-strong leaf typicality for the same setup as in Figure~\ref{f:ETHtime1}, with $\beta_0=0.75$.
\textsc{left:} The points represent the expectation values of the spin operators in the pure states of the minimum-variance ensemble for a chain of $L=16$ spins. The curves show the estimates of $f_{\sigma^\alpha,\mathcal L_H}(E)$ used to identify outliers.
\textsc{right:} For each spin operator on the left, the corresponding panel shows the normalized logarithm of the number of outliers as a function of the threshold $\Delta$ (see main text), for the system sizes shown in the legend; thicker lines correspond to larger sizes.
}\label{f:ETH1}
\end{figure}

\begin{figure}[t]
\centering
\includegraphics[width=0.7\textwidth]{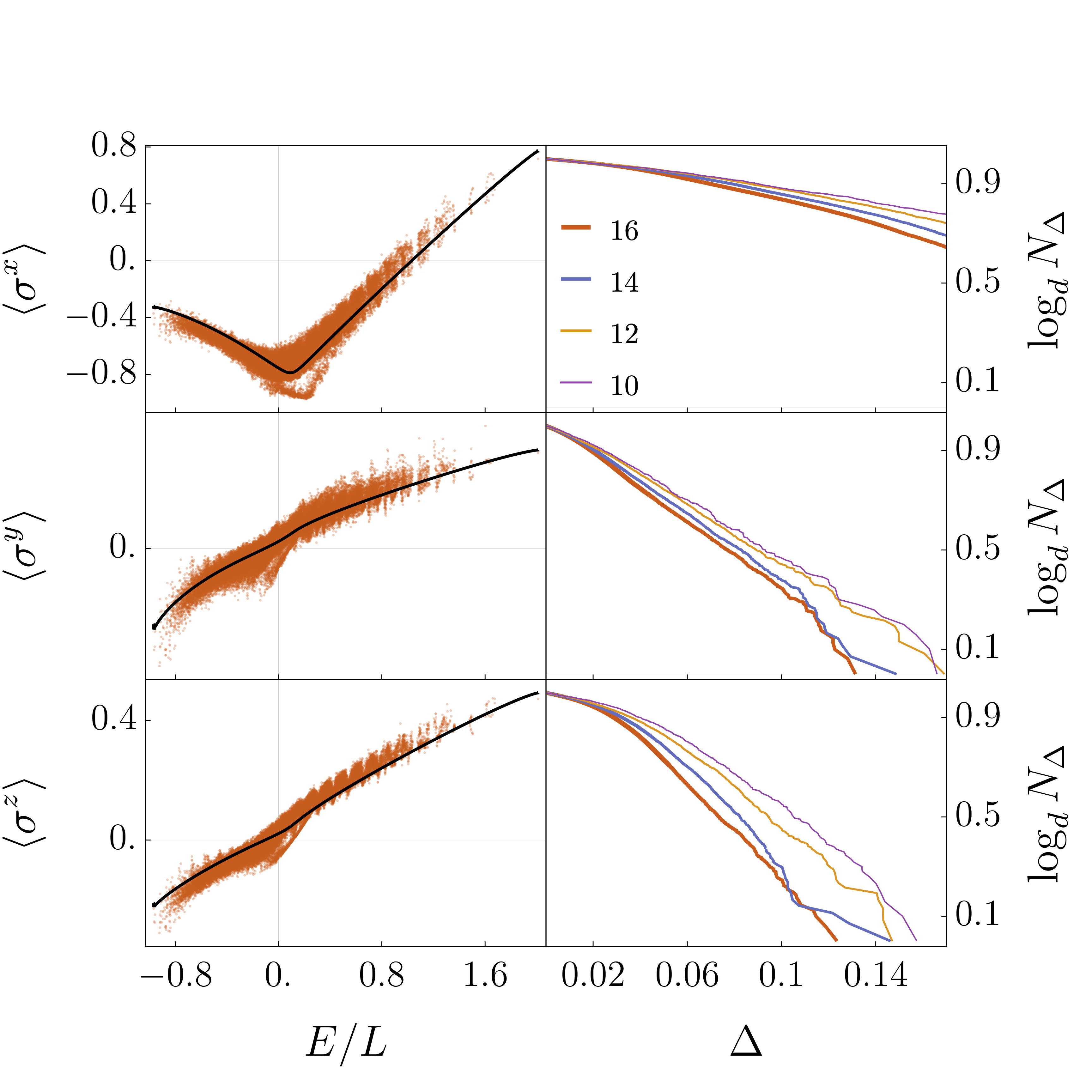}
\caption{
Indications of the absence of near-strong leaf typicality for the same setup as in Figure~\ref{f:ETH1}, but with the initial state rotated so that $H_0=\sum_{\ell}\sigma_\ell^z\sigma_{\ell+1}^z+1.5\sigma_\ell^x$.
\textsc{left:} A separate branch appears in the bulk of the energy spectrum, particularly visibly in the first panel.
\textsc{right:} As $L$ increases, the $\sigma^x$ curves appear to approach a nonzero limiting curve for small, fixed thresholds $\Delta>0$. 
}\label{f:ETH2}
\end{figure}

To reveal the strength of leaf typicality, Figure~\ref{f:ETH1} shows the local spin expectation values for the thermal state of $H_0$ at $\beta_0=0.75$, whose leaf is distinct from the commuting leaf of $H$.
The normalized logarithm of the number of states \(N_\Delta\) whose expectation value differs by more than \(\Delta\) from the typical value at the corresponding energy is plotted as a function of \(\Delta\). 
The trend is consistent with near-strong leaf typicality (Materials and Methods). 
Figure~\ref{f:ETH2} shows that this conclusion cannot be extended to every leaf of the foliation. Indeed, simply rotating the initial state leads to a leaf that appears to require a more elaborate thermodynamic compression. In particular, our diagnostic is consistent with weak leaf typicality but indicates a breakdown of near-strong leaf typicality when we insist on using a single limiting function $f_{O,\mathcal L_H}(E)$ to describe local properties across the entire leaf.

More generally, our numerical analysis of a variety of  leaves obtained by changing $H_0$ and $H$ 
has revealed two main obstructions to near-strong leaf typicality, which we present as motivations for further investigation: (i) there is an additional operator with quasilocal density, independent of $H_{\rho}$ and the identity, that commutes, or approximately commutes, with the dressed Hamiltonian, (ii) even in the zero temperature limit, the energy $\lim_{\beta_0\rightarrow\infty}\mathrm{tr}(\rho_{\beta_0}H)$ remains in the bulk of the spectrum of the corresponding dressed Hamiltonian $\lim_{\beta_0 \rightarrow\infty}H_{\rho_{\beta_0}}$. Ref.~\cite{Fagotti2026Temperature} provides a justification of obstruction~(i) in the thermodynamic limit. Qualitatively, it is analogous to the effect of additional conservation laws on ETH. Obstruction~(ii) is subtler and might be connected with other known counterexamples to ETH~\cite{Shiraishi2017Systematic}. Strictly speaking, condition~(ii) singles out one atypical representative on the limiting leaf as $\beta_0\rightarrow\infty$; however, in every case examined, this was accompanied by an exponentially growing number of atypical representatives also on leaves at finite $\beta_0$. The setting in Figure~\ref{f:ETH2} illustrates this second obstruction.
Remarkably, the expectation values shown in the left panels appear to form two branches, most clearly for $\sigma^x$.
This raises the question of whether the observed phenomenology could reflect a phase transition on the leaf in the thermodynamic limit.

\section*{Discussion}
We have proposed the minimum-variance foliation of quantum state space as an organizing structure for thermodynamics beyond equilibrium. Within this structure, we have formulated a leaf-typicality hypothesis that would give the leaves a thermodynamic role through the local equivalence of their isoenergetic pure-state representatives. 
This lays the foundation for extending the language of equilibrium statistical mechanics away from equilibrium, as well as for identifying thermodynamic phenomena that emerge only in the presence of energy coherence.

Many open questions remain. We deliberately restricted our analysis to the case where $H_\rho$ is nondegenerate. We expect accidental degeneracies of $H_\rho$ to leave the overall physical picture intact. Nevertheless, important classes of states outside $\mathcal M_H$ deserve special attention. These include pure states, which play a paradigmatic role in quench dynamics. While a pure state that is the ground state of a local Hamiltonian may be interpreted as the zero-temperature limit of a thermal state, such a procedure is ambiguous (many Hamiltonians share the same ground state), hence the same pure state generically lies at the intersection of the closures of a multitude of leaves. This points to the possibility that atypical states, such as leaf analogues of many-body scars, arise naturally at intersections of leaf closures.

A central direction is the formulation of the theory directly in the thermodynamic limit. This program has begun in Ref.~\cite{Fagotti2026Temperature}, which extends the notion of temperature beyond the commuting leaf and studies its behavior under restriction to subsystems. However, the dynamics of relaxation induced by subsystem maps remains an open question. 

Finally, we have not fully characterized the mechanisms responsible for the breakdown of near-strong leaf typicality, which might even go beyond the leaf adaptations of the known mechanisms undermining ETH.

\subsection*{Materials and Methods}

\subsubsection*{Thermodynamic entropy}
We show here that the thermodynamic entropy satisfies basic thermodynamic properties. First, consider an initial state $\rho$ satisfying $[\rho,H(0)]=0$, which evolves in a closed system under a time-dependent Hamiltonian $H(t)$. Its thermodynamic entropy cannot fall below its initial value: $S_{th}^{(H(t))}[\rho(t)] \geq S_{th}^{(H(0))}[\rho]$.
Indeed, the von Neumann entropy of a mixture of pure states is bounded above by the Shannon entropy of the mixing probabilities. Applying this bound to the optimal decomposition at time $t$ gives
$S_{vN}[\rho(t)] \leq S_{th}^{(H(t))}[\rho(t)]$.
The result then follows from the unitary invariance of the von Neumann entropy and its equality with the thermodynamic entropy for the initial stationary state. Analogously, for an initially uncorrelated pair of stationary subsystems $A$ and $B$, after they are put in contact, the sum of their thermodynamic entropies cannot fall below its initial value, indeed $S_{th,A}(t)+S_{th,B}(t)\geq S_{vN}[\rho_A(t)]+S_{vN}[\rho_B(t)]\geq S_{vN}[\rho_{AB}(t)]=S_{vN}[\rho_{AB}(0)]=S_{th,A}(0)+S_{th,B}(0)$. Other thermodynamic relations involving temperature cannot be discussed without a proper definition of the latter, for which we  refer the reader to Ref.~\cite{Fagotti2026Temperature}.  

\subsubsection*{Quantifying energy (in)coherence}
The coherent structure in the energy eigenbasis~\cite{Streltsov2017Colloquium,Lostaglio2015Description} is encoded in the variance-minimizing optimal family $\{|\varphi_i\rangle\}$, while varying the populations $\{p_i\}$ amounts to classical mixing within that structure.
The commuting leaf plays the role of an incoherent sector~\cite{Baumgratz2014Quantifying}, in which the optimal family coincides with the energy eigenbasis and is orthonormal.
Conversely, for a full-rank reference state, an orthonormal optimal family occurs only on the commuting leaf; thus, coherence across energy levels is faithfully witnessed by the nonorthogonality of $\{|\varphi_i\rangle\}$.
A natural leaf-level incoherence indicator is provided by the barycenter
$
\rho_\bullet(\mathcal L_H):=\frac{1}{d}\sum_{i=1}^{d}|\varphi_i\rangle\langle\varphi_i|
$, 
and its von Neumann entropy,
\begin{equation}\label{eq:incoherence}
\mathfrak I(\mathcal L_H)=S_{vN}[\rho_\bullet(\mathcal L_H)]=-\mathrm{tr}[\rho_\bullet(\mathcal L_H)\log\rho_\bullet(\mathcal L_H)]\, .
\end{equation}
This is reminiscent of quantifiers of ``ensemble quantumness'' based on overlaps~\cite{Fuchs2003Squeezing,Sun2021Quantumness,Theurer2017Resource}.
It satisfies $0\le \mathfrak I(\mathcal L_H)\le \log d$, and it is maximal, $\mathfrak I(\mathcal L_H)=\log d$, if and only if $\rho_\bullet(\mathcal L_H)=\mathrm{I}/d$, which occurs precisely on the commuting leaf.
At the opposite extreme, $\mathfrak I(\mathcal L_H)$ can become arbitrarily small when the optimal states $\{|\varphi_i\rangle\}$ become nearly collinear and $\rho_\bullet(\mathcal L_H)$ approaches a pure state.

\subsubsection*{Quantifying typicality}
To quantify leaf typicality, for each local observable $O$ we compute the expectation values $\langle\varphi_i|O|\varphi_i\rangle$ in the optimal ensemble. 
Rather than estimate $f_{O,\mathcal L_H}(E)$ from energy-resolved local averages, we exploit the regularity properties discussed in Ref.~\cite{Fagotti2026Temperature} to estimate it from the expectation value of $O$ in the isoenergetic leaf canonical ensemble through the initial state $\rho$, with density matrix proportional to $\sqrt{\rho}e^{-\Delta\beta H_\rho}\sqrt{\rho}$ and $\Delta\beta$ chosen to give mean energy $E$. We use the data only to center the residuals by subtracting their unweighted mean, which we interpret as a finite-size bias. 
We then define $N_\Delta$ as the total number of representatives $|\varphi_i\rangle$ for which
$|\langle\varphi_i|O|\varphi_i\rangle-f_{O,\mathcal L_H}(E_{i})|>\Delta$ and consider $\log_{d} N_\Delta$ as a function of $\Delta$. 
This diagnostic is closely related to large-deviation analyses of the fraction of athermal eigenstates in a microcanonical window~\cite{Yoshizawa2018Numerical,Kim2014Testing}.
It probes both weak and near-strong leaf typicality. For the former, it is sufficient that, for fixed $\Delta>0$,
$
\limsup_{L\to\infty}\log_{d} N_\Delta<1
$.
Note that a nonzero asymptotic limit signals an exponentially large number of atypical states in the optimal decomposition, as occurs for integrable systems on the commuting leaf~\cite{Mori2016Weak}. Near-strong typicality is more restrictive and corresponds here to
$
\lim_{L\to\infty}\log_{d} N_\Delta=0
$.

\subsubsection*{QFI and optimal energy variance}
The QFI with respect to \(H\) can be written as the variance of the symmetric logarithmic derivative \(L_\rho^{(H)}\), defined by
\begin{equation}\label{eq:Lyapunov}
\tfrac{1}{2}\{\rho,L_\rho^{(H)}\}=i[\rho,H]\, .
\end{equation}
We first show that all density matrices in a leaf have the same SLD. Using \eqref{eq:Hrho}, one has
\begin{equation}
L_\rho^{(H)}=i(H_\rho^- - H_\rho^+)\, ,\qquad 
H=\tfrac{1}{2}(H_\rho^-+H_\rho^+)\, ,
\end{equation}
where \(H_\rho^s=\rho^{-s/2}H_\rho\rho^{s/2}\). Since
\(H_\rho^-|\varphi_i\rangle=E_{i}|\varphi_i\rangle\), it follows that
\[
L_\rho^{(H)}|\varphi_i\rangle
=
2i(E_{i}-H)|\varphi_i\rangle,
\]
which does not depend on the populations \(\{p_i\}\). Since $\rho$ is full-rank, the set \(\{|\varphi_i\rangle\}\) forms a basis, and hence 
\begin{equation}
L_{\rho'}^{(H)}=L_\rho^{(H)} \qquad\text{for every}\quad \rho'\in\mathcal L_H(\rho)\, .
\end{equation}

Let us now restrict to quantum spin chains with sufficiently short-range interactions and consider the physical situation in which $\rho$ is a Gibbs ensemble of a local Hamiltonian $H_0$. In this setting, complex-time analyticity of local observables follows from the standard locality/analyticity results for quantum spin chains~\cite{Araki1969Gibbs}. In our specific situation, it implies that $L_\rho^{(H)}$ admits a quasilocal density.  Indeed, for Gibbs states of $H_0$ with inverse temperature $\beta_0, $ $H_\rho$ and $L^{(H)}_\rho$ can be represented as averages of $H(t)$, evolved with $H_0$, with kernels of typical width $\beta_0$~\cite{Fagotti2026Temperature}. If $h_\ell$ is a local operator representing the density of $H$ around site $\ell$, the corresponding quasilocal SLD density can be obtained as $L_\rho^{(h_\ell)}=\lim_{n\rightarrow\infty}L_{\rho_{A_n}}^{(h_\ell)}$, where $\rho_{A_n}$ is the reduced density matrix of the subsystem $A_n=\{\ell-n,\ell-n+1,\ldots,\ell+n\}$. Since this density is determined by the local properties of $\rho$, two states belonging to thermodynamically distinct leaves, characterized by inequivalent quasilocal SLD densities, cannot be locally equivalent.

The quasilocality of the SLD density also has immediate consequences for the QFI. Indeed, from \eqref{eq:Lyapunov} it follows
\begin{equation}\label{eq:EVFO}
F_Q(\rho;H)=i\,\operatorname{tr}(\rho[H,L_\rho^{(H)}]).
\end{equation}
Since commutation with a local Hamiltonian preserves quasilocality, \eqref{eq:EVFO} represents the QFI as the expectation value of an extensive observable with quasilocal density.
Leaf typicality applied to this quasilocal observable implies that typical isoenergetic representatives in the optimal ensemble of $\mathcal L_H(\rho)$ have the same QFI density, which, for pure representatives, is four times their energy-variance density. 

\section*{Acknowledgments}
I thank Florent Ferro and Luca Capizzi for discussions.

%\clearpage 

\bibliography{references}
\bibliographystyle{sciencemag}

\end{document}